\documentclass[twoside,twocolumn]{article}

\usepackage{blindtext} 
\usepackage{multicol}
\usepackage[T1]{fontenc} 
\usepackage[english]{babel} 
\usepackage[hmarginratio=1:1,top=28mm, bottom=30mm, columnsep=30pt, left=2cm, right=2cm]{geometry} 
\usepackage[hang, small,labelfont=bf,up,textfont=it,up]{caption} 
\usepackage{booktabs} 
\usepackage{enumitem} 
\setlist[itemize]{noitemsep} 

\usepackage{abstract} 

\usepackage{titlesec} 
\renewcommand\thesection{\Roman{section}} 
\titleformat{\section}[block]{\normalsize\bf}{\thesection.}{1em}{} 
\titleformat{\subsection}[block]{\normalsize}{\thesubsection.}{1em}{} 
\usepackage[utf8]{inputenc}
\usepackage{authblk}
\usepackage{mathrsfs}
\usepackage{amssymb}
\usepackage{amsmath}
\usepackage{graphicx}
\usepackage{mathptmx}
\usepackage{txfonts}
\usepackage{titling} 
\usepackage{hyperref} 

\pretitle{\begin{center}\huge\bfseries} 
\posttitle{\end{center}} 
\title{\Large A possible time dependent generalization of the bipartite quantum marginal problem} 
\author{Giuseppe Baio\textsuperscript{1}, Dariusz Chru\'{s}ci\'{n}ski\textsuperscript{2} \& Antonino Messina\textsuperscript{3,4}} 
\affil{\textsuperscript{1}\textit{\small SUPA and Department of Physics, University of Strathclyde, Glasgow G4 0NG, Scotland, U.K.},\\
\textsuperscript{2}\textit{\small Nicolaus Copernicus University,  Grudzi\k{a}dzka 5/7, 87–100 Toru\'{n}, Poland},\\
\textsuperscript{3}\textit{\small Dipartimento di Matematica e Informatica, Università degli Studi di Palermo, Italy, }\\
\textsuperscript{4}\textit{\small I.N.F.N., Sezione di Catania, Italy.}} 
\affil{\small*\textbf{Corresponding author}: giuseppe.baio@strath.ac.uk} 

\date{} 
\renewcommand{\maketitlehookd}{%
\begin{abstract}
In this work we study an inverse dynamical problem for a bipartite quantum system governed by the time local master equation: to find the class of generators which give rise to a certain time evolution with the constraint of fixed reduced states (marginals). The compatibility of such choice with a global unitary evolution is considered. For the non unitary case we propose a systematic method to reconstruct examples of master equations and address them to different physical scenarios.
\end{abstract}
}

\begin{document}
\maketitle

\section{Introduction}

The preparation on demand of a microscopic system in a quantum state of interest is nowadays regarded as a reachable target in several experimental contexts, such as control of chemical reactions, selective population transfers and nuclear magnetic resonance \cite{traincat}.
Manipulating the dynamics of a microscopic system can be thought as the ability to drive its time evolution along a specific path in its space of states \cite{schirm1}. However, an effective control of the dynamics of a quantum system, although still a far-reaching challenge, corresponds to a crucial step towards the realization of several quantum technological applications \cite{wiseman},\cite{altafini}.  

On a further complexity level, if we assume to deal with a \textit{bipartite} system A + B, we can suppose that in an experiment we have access only to the reduced states of the two subsystems. What properties of the global system can we deduce from such time dependent single party information only? Partial replies have been given to such a question concerning multipartite entanglement \cite{gross}, pure quantum states \cite{schilling1} and open quantum system dynamics \cite{mazziotti} but general results are still not available so far. 

In this paper we face this problem in the light of control for a bipartite quantum system.
We assume indeed to assign the time evolution of the reduced states of both subsystems $A$ and $B$. The question posed is how to construct a protocol aimed at engineering the realization of a global physical scenario evolving in accordance with the constraints imposed on both the reduced quantum dynamics \cite{schirm2}.

Mathematically speaking, in order to tackle such problem, one is reduced to the following question:	
to single out from a joint density matrix $\rho_{AB}(t)$ at least one possible \textit{generator} of its time evolution, namely a superoperator $\mathcal{L}_{t}$ such that: 

\begin{equation}
\dot{\rho}_{AB}(t)=\mathcal{L}_{t}[\rho_{AB}(t)].  
\label{gen}
\end{equation}
 
The structure of the generator $\mathcal{L}_{t}$ then contains all the information about the system, whether it is isolated or interacting with an environment \cite{BP}. 
 
If we restrict ourselves to unitary dynamics only, we simply have that $\mathcal{L}_{t}=-i\left[H_{AB}(t),\cdot\:\right]$ so that our problem can be addressed to the reconstruction of the class of self-adjoint bipartite Hamiltonians $H_{AB}(t)$ from a certain $\rho_{AB}(t)$ compatible with the assigned reduced density matrices $\rho_{A}(t)$ and $\rho_{B}(t)$. 
In ref. \cite{bernatska}, a method of reconstruction of Hamiltonians from a given time evolution was provided by means of a \textit{stereographic parametrization} of unitary operators. Such a method is based on the \textit{foliation} of the space of states into non intersecting  \textit{orbits} \cite{bernatska2}.

Generalizing the inverse dynamical problem posed here for a dissipative evolution, i.e. for an \textit{open} quantum system, implies many complications. Firstly, we have a much larger freedom due to the lack of knowledge about the total system and, most importantly, necessary and sufficient conditions on the superoperator $\mathcal{L}_{t}$ of eq. (\ref{gen}) to be physically legitimate are generally not known \cite{Chru1}.  Nevertheless, special classes of valid generators can be identified and their mathematical structure reflect important properties of the environment \cite{rivas}. 

A second fundamental implication is given by the constraints on the global state evolution $\rho_{AB}(t)$ imposed by the knowledge of its reduced density matrices $\rho_{A}(t)$ and $\rho_{B}(t)$ or \textit{marginals}. The search of such constraints shows a connection with the \textit{Quantum Marginal Problem} (QMP) which is of interest in several contexts, ranging from quantum chemistry to quantum information theory \cite{marginal},\cite{schilling},\cite{kuskus}. In our framework, they appear naturally as an aspect regarding \textit{kinematics} providing thus a first time-dependent generalisation of the QMP. 
Moreover, assuming knowledge of a joint state $\tilde{\rho}_{AB}(t)$ from the answer to the above problem, a subsequent \textit{dynamic} QMP generalization can be formulated: to tailor the features of a physical system dynamically following the trajectory $\tilde{\rho}_{AB}(t)$ from $\tilde{\rho}_{AB}(t_{0})$ and then satisfying:  

\begin{equation}
\rho_{A}(t)=\textrm{Tr}_{B}\tilde{\rho}_{AB}(t) \quad \rho_{B}(t)=\textrm{Tr}_{A}\tilde{\rho}_{AB}(t)
\label{traces}
\end{equation}

at any time instant \footnote{The notation $\textrm{Tr}_{B}$ indicates the \textit{partial trace} i.e.  trace over states in $\mathcal{H}_{B}$}.
In this work we set the stage by analyzing such problems for the lowest dimensional case, i.e. those of two interacting qubits, showing how such apparently simple cases already present a remarkable physical and mathematical richness. 
The paper is organized as follows. In section \ref{2}, we recall the mathematical notation and the fundamental formalism of closed and open system quantum dynamics. In section \ref{3} and \ref{4}, we introduce instead the two classes of problems which can be thought as possible generalizations of the QMP to the time dependent domain. The method described in each section is applied to specific examples of physical interest.

\section{Time evolution of quantum systems and local generators}
\label{2}

Before formulating the problems let us provide a concise introduction to closed and open quantum dynamics. 
Firstly, the space of density matrices of a quantum system whose Hilbert space $\mathcal{H}$ has dimension $n$ can be identified with the convex positive definite subset of the affine linear space:

\begin{equation}
\mathcal{S}_{n}=i\mathbf{su}(n)\oplus\left(\frac{1}{n}\right)\hat{\mathbb{I}}_{n}.
\label{lieAB}
\end{equation}

where $\mathbf{su}(n)$ is the Lie algebra of the special unitary group \textit{SU}($n$), namely the space of skew-Hermitian matrices \cite{bernatska}. 
Any time evolution for a closed quantum system satisfies the \textit{Liouville-Von Neumann} equation:

\begin{equation}
\dot{\rho}(t)=-i\left[H(t),\rho(t)\right],
\label{vonneum}
\end{equation}

where $[A,B]=AB-BA$ and the self-adjoint operator $H(t)$ is the \textit{Hamiltonian}. The evolution generated by a self-adjoint operator $H(t)$ can always be represented by a unitary operator so that the solution eq. (\ref{vonneum})  reads:

\begin{equation}
\rho(t)=U_{t,t_{0}}\rho(t_{0})U^{\dag}_{t,t_{0}}.
\label{urhou}
\end{equation}

In a geometrical language, one says that the space of states of a quantum system is \textit{foliated} by the action of the unitary group into a set of non-intersecting \textit{orbits} labeled by the spectrum of $\rho(t_{0})$ \cite{bernatska2}. 
Moreover, orbits are uniquely determined by a set of parameters called \textit{trace invariants} \cite{messina1}. This characterization is provided by the following preposition:\newline

\textbf{Preposition 1}: \textit{A density matrix $\rho(t)$ evolves unitarily if and only if:}

\begin{equation}
\textrm{Tr}[\rho^{k}(t)]=\textrm{Tr}[\rho^{k}(t_{0})]=\textrm{const}.
\label{tracesk}
\end{equation}

\textit{where $1\leq k \leq n$}.\newline

\textit{Proof} - If $\rho(t)$ evolves unitarily, eq.(\ref{tracesk}) holds because the trace operation is invariant under every similarity transformation. 
On the other hand, denoting with $\lambda_{i}(t)$ the eigenvalues of $\rho(t)$ and assuming all the matrix powers of $\rho(t)$ constants in time, we have that the characteristic polynomial can be expanded in a power series with coefficients given by the elementary symmetric polynomials $e_{i}$, functions of the roots $\lambda_{i}(t)$ \cite{macdonald}:

\begin{equation}
\prod^{n}_{i=1}\left[\lambda-\lambda_{i}(t)\right] = \sum^{n}_{k=0}(-1)^{n+k}e_{n-k} \lambda^{k}
\label{poli}
\end{equation}

Each $e_{i}$ appearing in the left side of eq. (\ref{poli}) can be related through the \textit{Newton-Girard identities} to the $k$-th power sum $p_{k}$, thus:

\begin{equation}
p_{k}\equiv\sum^{n}_{i=1}\lambda^{k}_{i}(t)=\textrm{Tr}(\rho(t)^{k}) = \textrm{const}.
\end{equation}

Therefore, the coefficients of the characteristic polynomial of $\rho(t)$ are functions on the $p_{k}$ only which, by assumption, are constants and so do the eigenvalues $\lambda_{i}(t)$. \newline

Generally speaking, the evolution of a quantum system denoted with $S$ coupled to an environment denoted with $E$, is obtained performing the partial trace over a larger joint quantum state $\rho_{SE}(t)$, namely:

\begin{equation}
\rho(t)=\textrm{Tr}_{E}\left[U_{t,t_{0}}\rho_{SE}(t_{0})U^{\dag}_{t,t_{0}}\right],
\label{partial}
\end{equation}

where $\rho_{SE}(t_0)$ is the initial joint state. 

If we assume the two systems being in a product state at time $t=t_{0}$, i.e. $\rho_{SE}(t_{0})=\rho(t_{0})\otimes\rho_{E}(t_{0})$, eq. (\ref{partial}) defines, at any time instant, a one parameter family of maps from the space of states of system $S$ onto itself and can be represented as follows: 

\begin{equation}
\rho(t)=\Lambda_{t,t_{0}}\left[\rho(t_{0})\right].
\label{map} 
\end{equation}

Moreover, the map $\Lambda_{t,t_{0}}$ defined by this prescription satisfies always the property of being \textit{completely positive} (CP) and \textit{trace preserving} (TP) and therefore called CPTP or \textit{dynamical map} \cite{alicki2}.
If the map is invertible at any $t$, the generator appearing in eq. (\ref{gen}) is simply given by: 

\begin{equation}
\mathcal{L}_{t}=\dot{\Lambda}_{t,t_{0}}\Lambda^{-1}_{t,t_{0}},
\label{convolutionless} 
\end{equation}

or, equivalently, the dynamical equation for the map can be written in the  time convolutionless form $\dot{\Lambda}_{t}=\mathcal{L}_{t}\Lambda_{t}$. Such description gives rise to the so-called \textit{time local master equation}. 
In the case when the generator is time independent, the map $\Lambda_{t}$ fulfills the semigroup property \footnote{$\Lambda_{t_{1}}\Lambda_{t_{2}}=\Lambda_{t_{1}+t_{2}}$ for any $t_{1},t_{2}\geq0$} and we recover the result of Gorini, Kossakowski and Sudarshan \cite{GKS} and Lindblad \cite{Lindblad}, namely that the generator of a quantum dynamical semigroup has the following form (GKSL):

\begin{equation}
\dot{\rho}(t)=-i\left[H,\rho(t)\right]+\sum_{k}\left(V_{k}\rho(t) V_{k}^\dag -\frac{1}{2}\{V_{k}^\dag V_{k},\rho(t)\}\right).
\label{GKSL}
\end{equation}

where  $V_{k}$ are operators in $\mathcal{H}$ and $\{A,B\}=AB+BA$.

Relaxing the assumption of a time independent generator leads to an uncharted territory. In other words, necessary and sufficient conditions on $\mathcal{L}_{t}$ guaranteeing that the corresponding dynamical map is CP are not known. To our knowledge, the only class of generators which was partially characterized corresponds to the time commutative case, i.e. $[\mathcal{L}_{t},\mathcal{L}_{u}]=0$ for any $t,u \geq 0$. \cite{Chru1}. From the next section we start discussing the time dependent problems anticipated above based on the reduced state picture. 

\section{Time dependent marginals}
\label{3}

We can now discuss our class of problems for two qubits $A$ and $B$. We distinguish between a time dependent QMP in a \textit{kinematic} and \textit{dynamic} sense but the two clearly originate from the following:\newline

\textbf{Problem}: \textit{Given a bipartite system $A+B$, assign at will a couple of reduced evolutions of interest, $\rho_{A}(t)$ and $\rho_{B}(t)$. Characterize the features of the global time evolution $\rho_{AB}(t)$ that can be inferred from the knowledge of such constraints.} \newline


Suppose we are given two time dependent marginals, denoted with $\rho_{A}(t)$ and $\rho_{B}(t)$. We ask first whether it is possible in principle to obtain them from a certain joint state $\rho_{AB}(t)$, i.e a density matrix such that eq. (\ref{traces}) holds at every time instant.
The purpose of this section is then to classify all states $\rho_{AB}(t)$ with the assigned marginals and then single out in this class examples which can be dealt with as possible real physical scenarios.
We choose the term Kinematic QMP to indicate what follows:\newline

\textbf{Kinematic QMP:} \textit{Given at will two time dependent marginals $\rho_{A}(t)$ and $\rho_{B}(t)$, characterize the class of global states $\rho_{AB}(t)$ compatible with $\rho_{A}(t)$ and $\rho_{B}(t)\,\, \forall t$.} \newline

Let us start noticing that there exists at least one trivial answer of the above problem, namely the tensor product $\rho_A(t)\otimes\rho_B(t)$. For all other cases, we adopt the decomposition of $\rho_{AB}(t)$ with respect to eq. (\ref{lieAB}) that gives rise to the generalization of the well known Bloch vector parametrization when $n=2$. In the case of two qubits, a generic element of $\mathcal{S}_{2\times2}$ reads as follows (we omit time dependence) \cite{fano}:

\begin{equation}\begin{split}
\rho_{AB}=\frac{1}{4}\left(\mathbb{I}_{4}+\sum^{3}_{i=1}x_{i}\sigma^{i}_{A}\otimes\mathbb{I}_{B}+\sum^{3}_{j=1}y_{j}\mathbb{I}_{A}\otimes\sigma^{j}_{B} \right. \\
\left. + \sum^{3}_{i,j=1}z_{ij}\sigma^{i}_{A}\otimes\sigma^{j}_{B}\right), 
\label{general}
\end{split}\end{equation}

where the 15 real coefficients represent the \textit{coherence vector} $\textbf{r}=[x_{i},y_{i},z_{ij}]^\textrm{T}$ and $\sigma^{i}\,\,i=1,2,3$ are the Pauli matrices. A joint density matrix  $\rho_{AB}$ can thus be simply expressed exploiting the following suitable representation:

\begin{equation}
\rho_{AB}=\rho_{A}\otimes\rho_{B}+\Delta_{\textrm{corr}}.
\label{corre}
\end{equation}

Therefore, the class of $\rho_{AB}$  with fixed partial traces $\rho_{A}$ and $\rho_B$ is given by all 9 possible entries of the \textit{correlation} tensor, namely:

\begin{equation}
\Delta_{\textrm{corr.}}=\frac{1}{4}\sum_{i,j}\tilde{z}_{ij}\sigma^{i}_{A}\otimes\sigma^{j}_{B},
\label{corrtransf}
\end{equation}

where $\tilde{z}_{ij}=z_{ij}-x_{i}y_{j}$ and $\Delta_{\textrm{corr}}$ has vanishing partial traces, i.e. $\textrm{Tr}_{A}\Delta_{\textrm{corr}}=\textrm{Tr}_{B}\Delta_{\textrm{corr}}=0$. Therefore, the answer to the kinematic QMP is given by all possible time dependent $\Delta_{\textrm{corr}}(t)$ such that the form (\ref{corre}) is positive definite.
The main difficulty concerning the above parametrization is that, in order to construct examples within such class, the conditions ensuring positive semi-definiteness at any time instant of $\rho_{AB}(t)$ can be unpractical for $n\geq3$ \cite{messina1},\cite{kimura1}.  
However, one can show that in this case $\Delta_{\textrm{corr.}}$ can always be brought by two independent \textit{local} transformations in the following form \cite{huo}: 

\begin{equation}\begin{split}
&\Delta_{\textrm{corr.}}=\frac{1}{4}\sum_{i}\tilde{z}_{ii}\sigma^{i}_{A}\otimes\sigma^{i}_{B} = \\
&\left(
\begin{array}{cccc}
 \tilde{z}_{33} & 0 & 0 & \tilde{z}_{11}-\tilde{z}_{22} \\
 0 & -\tilde{z}_{33} & \tilde{z}_{11}+\tilde{z}_{22} & 0 \\
 0 & \tilde{z}_{11}+\tilde{z}_{22} & -\tilde{z}_{33} & 0 \\
 \tilde{z}_{11}-\tilde{z}_{22} & 0 & 0 & \tilde{z}_{33} \\
\end{array}
\right).
\label{formX}
\end{split}\end{equation}

We will use such restricted class to show whether it is possible or not to find a unitarily evolving $\rho_{AB}(t)$ within the solution of a certain kinematic QMP, i.e. all compatible states. Indeed, if an attempt of constructing such $\rho_{AB}(t)$ within this class fails, one can safely conclude that there are no other counterparts in the general class.  For any other $\rho_{AB}(t)$ of the form (\ref{general}) there are general approaches to ensure $\rho_{AB}\geq 0$,  e.g. inequalitities based on Newton-Girard identities \cite{kimura1} or  matrix factorization theorems \cite{matrix}.

%

In what follows we provide two of examples to show how fixed time dependent marginals imply constraints on the global time evolution. The next section is devoted to characterize possible physical scenarios from which the two marginals originate. We classify such an inverse problem as a \textit{dynamic} quantum marginal problem, precisely defined in sec. \ref{4}.

\subsection{\textit{Example 1: global unitary evolution}}

Let us consider the following two simple marginals in diagonal form:

\begin{equation}\begin{split}
\rho_{A}(t)=\frac{1}{16}\left(\begin{array}{cc}
8-\cos(Jt)&0\\
0&8+\cos(Jt)
\end{array}\right),\\
\\
\rho_{B}(t)=\frac{1}{16}\left(\begin{array}{cc}
8+\cos(Jt)&0\\
0&8-\cos(Jt)\end{array}\right).
\end{split}\end{equation}

where $J>0$. It is easy to prove that the most general matrix form of eq. (\ref{general}) is the following: 

\begin{equation}\begin{split}
\rho_{AB}=\rho_{A}\otimes\rho_{B}+
\left(\begin{array}{cccc}
\epsilon&\Delta_{12}&\Delta_{13}&\Delta_{14}\\
&-\epsilon&\Delta_{23}&-\Delta_{13}\\
&&-\epsilon&-\Delta_{12}\\
\textrm{(c.c)}&&&\epsilon\end{array}\right),
\label{formgen}
\end{split}\end{equation}

where we denoted for simplicity with $\Delta_{12},\Delta_{13}, \Delta_{14}, \Delta_{23}$ the off diagonal coherences resulting from eq. (\ref{general}) and with $\epsilon$ the real parameter in the populations of $\rho_{AB}$. Once a legitimate value for $\epsilon$ is chosen, a simple approach to ensure $\rho_{AB}\geq 0$ is the following, based on \textit{Cholesky factorization}: we impose that $\rho_{AB}$
in eq. (\ref{formgen}) is factorizable in a certain matrix product i.e. $\rho_{AB}=LL^{\dag}$. In particular, $L$ must be a lower triangular matrix with non negative diagonal entries $L_{ii}$. Such entries can be calculated as functions of $\Delta_{12},\Delta_{13}, \Delta_{14}, \Delta_{23}$ and imposing $L_{ii}>0$ yields necessary and sufficient conditions for $\rho_{AB}\geq 0$. Therefore, the class of states singled out here provides the complete answer to this kinematic QMP.

In what follows we show how to achieve a unitarily evolving solution choosing the following matrix structure: 

\begin{equation}
\rho_{AB}(t) = \rho_A(t)\otimes\rho_B(t) + \left(
\begin{array}{cccc}
 \epsilon(t) & \cdot & \cdot & \cdot \\
 \cdot &  -\epsilon(t)& \Delta_{23}(t) & \cdot \\
 \cdot & \Delta^*_{23}(t) & -\epsilon(t)& \cdot \\
 \cdot & \cdot & \cdot & \epsilon(t) \\
\end{array}
\right).
\label{firstexample}
\end{equation}

\normalsize

with only one coherence $\Delta_{23}(t)\neq 0$.

It is easy to see that  imposing $\epsilon(t)=\frac{1}{256}\cos^2(Jt)$, each diagonal element of eq. (\ref{firstexample}) is positive and the positive semi-definiteness condition for $\Delta_{23}(t)$ becomes:

\begin{equation}
|\Delta_{23}(t)|^2\leq\frac{1}{256}(16-\cos^2(Jt)).
\end{equation}

According to eq. (\ref{tracesk}), a unitarily evolving $\rho_{AB}(t)$ would require the following trace invariants: 

\begin{equation}\begin{split}
\textrm{Tr}[\rho^2(t)] - \frac{1}{4} = 2 |\Delta_{23}(t)|^2+ \frac{1+\cos (2 J t)}{256} = c_1 \\
\textrm{Tr}[\rho^3(t)] -\frac{1}{16} =  \frac{3}{2} |\Delta_{23}(t)|^2+\frac{3 \cos ^2(J t)}{512} = c_2.
\label{rel1}
\end{split}\end{equation}

where $c_1$ and $c_2$ are two constants. Solving eq. (\ref{rel1}) for $|\Delta_{23}(t)|^2$ we get the following relation: 

\begin{equation}
c_1 -\frac{2}{3}c_2 = \cos(2Jt)- 2\cos^2(Jt) = -1.
\end{equation} 

Thus, since the LHS is time independent, it is possible to achieve a unitarily evolving $\rho_{AB}(t)$ with only $\Delta_{23}(t) \neq 0$. For example, choosing $c_1 = \frac{1}{128}$ we get: 

\begin{equation}
|\Delta_{23}|^2=\frac{\sin^2(Jt)}{256}
\end{equation} 

so that we can choose $\Delta_{23}=-i\frac{\sin(Jt)}{16}$, obtaining finally the following mixed and unitarily evolving density matrix:

\begin{equation}
\rho_{AB}(t)=\left(
\begin{array}{cccc}
 \frac{1}{4} & \cdot & \cdot & \cdot \\
 \cdot & \frac{1}{16} (\cos (J t)+4) & -\frac{1}{16} i \sin (J t) & \cdot \\
 \cdot & \frac{1}{16} i \sin (J t) & \frac{1}{16} (4-\cos (J t)) & \cdot \\
 \cdot & \cdot & \cdot & \frac{1}{4} \\
\end{array}
\right).
\label{nonuni}
\end{equation}

In sec. \ref{4} we show how to apply our procedure of reconstruction of a bipartite Hamiltonian of the composed system to this particular example.

\subsection{\textit{Example 2: global non-unitary evolution}} 

We show here that for an arbitrary choice of marginals a unitarily evolving joint state $\rho_{AB}(t)$ could not exist and thus our two-qubits system has to be open. In order to illustrate that, let us consider the following marginals:

\begin{equation}\begin{split}
\rho_{A}(t)=\frac{1}{2}\left(
\begin{array}{cc}
 1 & \cos (2 \omega t) \\
 \cos (2 \omega t) & 1 \\
\end{array}
\right),\\
\\
\quad \rho_{B}(t)=\frac{1}{2} \left(
\begin{array}{cc}
 1 & \sin (2\omega t) \\
 \sin (2\omega t) & 1 \\
\end{array}
\right),
\end{split}\end{equation}

where $\omega >0$. Again, the eigenbasis is time independent but the same is not true for the eigenvalues:


\begin{equation} \begin{split}
\textrm{spec}_{A}(t)=\left\{\alpha_1(t),\alpha_2(t)\right\} = \left\{\frac{1-\cos (2 t \omega )}{2},\frac{1+\cos (2 t \omega )}{2} \right\} \\
\textrm{spec}_{B}(t)=\left\{\beta_1(t),\beta_2(t)\right\} =\left\{\frac{1-\sin (2 t \omega )}{2},\frac{1+\sin (2 t \omega )}{2} \right\}, 
\end{split}\end{equation}

where $\textrm{spec}_{A}(t)$ indicates the spectrum of $\rho_{A}(t)$.
Thus, the two marginals are not isospectral (except in a countable set of times) and this leads us to conclude that a pure joint $\rho_{AB}(t)$ does not exist since, from the Schmidt decomposition, any pure state $|\Psi_{AB}(t)\rangle$ has isospectral marginals. Moreover, they are not even compatible with a unitarily evolving $\rho_{AB}(t)$. Indeed, denoting the two non-vanishing coherences with $\Delta_{14}(t)$ and $\Delta_{23}(t)$ , we choose: 

\begin{equation}
\rho_{AB}(t)=\left(\begin{array}{cccc}
\rho_{11}(t) &\cdot&\cdot&\Delta_{14}(t)\\
\cdot&\rho_{22}(t) &\Delta_{23}(t)&\cdot\\
\cdot&\Delta^{*}_{23}(t)&\rho_{33}(t)&\cdot\\
\Delta^{*}_{14}(t)&\cdot&\cdot&\rho_{44}(t)
\end{array}\right).\label{twocohe}\end{equation}

in the eigenbasis of the marginals. Necessary and sufficient conditions to achieve a unitarily evolving joint state $\rho_{AB}(t)$ equivalent to eq. (\ref{tracesk}) are the following \footnote{Such conditions arise from the characteristic polynomial of $\rho_{AB}(t)$.}:

\begin{equation}\begin{array}{c}
\rho_{11}(t)+\rho_{44}(t) = c\in [0,1]\\
\rho_{11}(t)\rho_{44}(t)-\left|\Delta_{14}(t)\right|^{2}= d_{1} \geq 0\\
\rho_{22}(t)\rho_{33}(t)-\left|\Delta_{23}(t)\right|^{2}= d_{2} \geq 0. 
\end{array}
\label{conduni2314}
\end{equation}

Recalling that $\rho_{11}(t)=\alpha_{1}(t)\beta_{1}(t)+\epsilon(t)$ and $\rho_{44}(t)=\alpha_{2}(t)\beta_{2}(t)+\epsilon(t)$, one must have: 

\begin{equation}
2\epsilon(t)=c-(\alpha_{1}(t)\beta_{1}(t)+\alpha_{2}(t)\beta_{2}(t)).
\end{equation}

On the other hand, the real number $0\leq c\leq 1$ must be such that the corresponding $\epsilon(t)$ keeps the populations of $\rho_{AB}(t)$ non-negative. It is easy to demonstrate that such condition on $c$ can be re-arranged in the following form: 

\begin{equation}\begin{split}
\max\left|\alpha_{1}(t)\beta_{1}(t)-\alpha_{2}(t)\beta_{2}(t)\right| \leq c  \\
\min\left[1-\left|\alpha_{1}(t)\beta_{2}(t)-\alpha_{2}(t)\beta_{1}(t)\right|\right] \geq c,
\label{conduni2314bis}
\end{split}\end{equation}

For the present example, such constant does not exist, since:

\begin{equation}\begin{array}{c}
\max\left|\alpha_{1}(t)\beta_{1}(t)-\alpha_{2}(t)\beta_{2}(t)\right|=\frac{1}{\sqrt{2}}\\
\min\left[1-\left|\alpha_{1}(t)\beta_{2}(t)-\alpha_{2}(t)\beta_{1}(t)\right|\right]=1-\frac{1}{\sqrt{2}}.
\end{array}\end{equation}

Therefore, we see in already simple cases that choosing time dependent correlations which ensure both positivity of the joint state $\rho_{AB}(t)$ and unitarity of its time evolution could lead to contradictions. As a general procedure one has to intersect contraints for positivity with those given by trace invariants of $\rho_{AB}^{k}(t),\,\, k=2,3$.\footnote{Only the first n-1 equation are \textit{functionally independent} for an n-level system. \cite{bernatska}}. In our two qubit case, with respect to the Bloch representation, one can derive the following two conditions: 

\begin{equation} \begin{split}
&I_{1} = \frac{1}{4}\left[\textrm{Tr}(\rho^2_{AB}(t))-\frac{1}{4}\right]  = |\mathbf{x}(t)|^2 + |\mathbf{y}(t)|^2 + \sum^{3}_{i=1} z^2_{ii}(t) = \textrm{const.} \\
&I_{2} =  \frac{1}{8}\left[\frac{1}{3}\left[\textrm{Tr}(\rho^3_{AB}(t))-\frac{1}{16}\right] -I_{1}\right] = \\
&\quad\quad\quad\quad\quad  \sum^{3}_{i=1}\left[ x_{i}(t)y_{i}(t)z_{ii}(t)\right] - z_{11}(t)z_{22}(t)z_{33}(t) = \textrm{const.}
\end{split}\end{equation}

Note that $|\mathbf{x}(t)| = x^2_{1}(t)+x^2_{2}(t)+x^2_{3}(t)$, i.e. the square modulus of the Bloch vector of system A. For an arbitrary choice of marginals, i. e. the two Bloch vectors $\mathbf{x}(t)$ and $\mathbf{y}(t)$, such an intersection may not exist.

\section{Dynamic Quantum Marginal Problem}
\label{4}

This last section focuses on our second class of time dependent marginal problems, namely the \textit{dynamic} QMP. Recalling the definition given above, the general target of a \textit{kinematic} QMP is to parametrize the class of all states $\rho_{AB}(t)$ whose time evolution realizes two marginals of interest. A dynamic counterpart of such problem can be stated as follows:\newline

\textbf{Dynamic QMP:} \textit{Given a joint state $\rho_{AB}(t)$ such that  the partial traces reproduce the two given marginals $\rho_A(t)$ and $\rho_{B}(t)$, characterize the class of time evolution generators.} \newline

Firstly, we discuss the case of unitarily evolving states. As said before, the problem of characterizing the physical scenario of the two qubits amounts essentially at the construction of a bipartite Hamiltonian $H_{AB}(t)$ which, from the initial state $\rho_{AB}(t_{0})$, generates the desired evolution with the fixed time dependent marginals. Finally, we provide also a particular protocol for constructing the form of a possible Master Equation for the case of dissipative evolution. 

\subsection{\textit{Unitary evolution: stereographic parametrization}}

Each unitary time evolution is realized within an orbit, namely the set of all states obtained by eq. (\ref{urhou}) when the evolution operator $U_{t,t_{0}}$ runs the group $\textrm{SU}(n)$. The diagonal form of $\rho(t_{0})=\Gamma$ identifies the orbit where $\rho(t)$ lies. 

For arbitrary system size, different $\rho(t_{0})$ lead to orbits of different dimensions. In particular, one has dimension $n(n-1)$ when $\Gamma=\textrm{diag}(\gamma_{1},\dots,\gamma_{n}),\, \gamma_{i} \in \left]0,1\right[$. Degenerate orbits are obtained when some $\gamma_{i}$ are equal. In the case when all $\gamma_{i}$ are equal but one (pure states), we have an orbit of dimension $2(n-1)$ \cite{bernatska}. Our target is then to achieve a parametrization of an \textit{evolution} matrix $\hat{u} \in \textit{SU}(n)$, such that:

\begin{equation}
\rho(t)=\hat{u}_{t}\Gamma\hat{u}_{t}^\dag.
\label{BM1}
\end{equation}

From group theory, each element of $\textrm{SU(n)}$ can be decomposed according the Gauss and Iwasawa decompositions for its complexified group $\textrm{SL}(n,\mathbb{C})$ \footnote{Namely the \textit{special linear} group of complex matrices with determinant equal to 1}\cite{hall2}. Indeed, defining an orbit as a coset space, in case of $\textrm{SU}(n)$, we have from Gauss decomposition that such space is parametrized by the subgroup of $\textrm{SL}(n,\mathbb{C})$ of complex lower triangular matrices, that we denote with $Z$. On the other hand, from Iwasawa decomposition we get that each $\hat{z}$ of $Z$ can be written in terms of a $\hat{u} \in \textrm{SU}(n)$ as:

\begin{equation}
\hat{z}=\hat{u}\hat{a}\hat{r},
\label{iwasawa}
\end{equation}

where $\hat{a}$ is an element of $A=\textrm{diag}(d_{1},\dots,{d_{n}}), \prod_{k}d_{k}=1$, (abelian subgroup of $\textrm{SL}(n,\mathbb{C})$) and $\hat{r} \in R$ which is a group of lower triangular matrices. One thus simply relates the entries of $\hat{a}$ and $\hat{r}$ to $\hat{z}$ as follows:

\begin{equation}
\hat{z}^*\hat{z}=\hat{r}^*\hat{a}^2\hat{r}
\label{zzrar}.
\end{equation}

Such an identification allows us to obtain a generic form of the operator $\hat{u}$ in terms of   $\hat{z}$ which parametrize the orbit and are called \textit{canonical coordinates}. Finally, the evolution operator $\hat{u}$ is calculated according to :

\begin{equation}
\hat{u}=\hat{z}\hat{r}^{-1}\hat{a}^{-1}.
\end{equation}

As a remark, the above parametrization methods yields an expression realizing a trajectory starting from the \textit{zero point} of the orbit, i.e. the point where all $\{z_{i}\}$ vanish. In order to obtain the \textit{actual} evolution operator, we multiply the parametrized $\hat{u}$ by the time dependent unitary operator which shifts the state from $\Gamma$ to $\rho(t_{0})$, namely:

\begin{equation}
\hat{U}_{t,t_{0}}=\hat{u}_t\hat{u}_{t_0}^{-1}.
\end{equation}

Nevertheless, it is easy to check that the Hamiltonian generating such evolution does not depend on this feature. Indeed one has:

\begin{equation}
\hat{H}(t)=i\left(\frac{d}{dt}\hat{U}_{t,t_{0}}\right)\hat{U}_{t,t_{0}}^{-1} = i\left(\frac{d}{dt}\hat{u}_t\right)\hat{u}_t^{-1}.
\label{recoham}
\end{equation}

For the two-qubit case, eq. (\ref{recoham}) admits an additive expansion in the basis of Hermitian operators in $\mathcal{H}_{A}\otimes\mathcal{H}_{B}$:

\begin{equation}
H_{AB}(t)=\sum_{\alpha,\beta=0}^{3}h_{\alpha\beta}(t)(\sigma^{A}_{\alpha}\otimes\sigma^{B}_{\beta}),
\label{fullham} 
\end{equation}
   
where $\sigma_{0}=\mathbb{I}$ and $h_{\alpha\beta}(t)=\textrm{Tr}(H_{AB}(t)\sigma_{\alpha}\otimes\sigma_{\beta})$. Therefore, eq. (\ref{fullham}) generally contains a \textit{local} term, i.e.

\begin{equation}
H_{A}(t)\oplus H_{B}(t)=\sum_{i=1}^{3}h_{i0}(t)(\sigma^{A}_{i}\otimes\mathbb{I}_{B})+\sum_{j=1}^{3}h_{0j}(t)(\mathbb{I}_{A}\otimes\sigma^{B}_{j}), 
\end{equation}

plus a remaining part $H_{AB}-H_{A}\oplus H_{B}$ can be defined as the interaction term $H_{I}$. Therefore, for the case of global unitary evolution, a full reconstructed Hamiltonian in the form of eq. (\ref{fullham}) completes the picture and the stereographic parametrization method provides the answer to any dynamical QMP for a unitarily evolving $\rho_{AB}(t)$. Let us apply such method to the example constructed in eq.(\ref{nonuni}), namely:

\begin{equation}
\rho_{AB}(t)=\left(
\begin{array}{cccc}
 \frac{1}{4} & \cdot & \cdot & \cdot \\
 \cdot & \frac{1}{16} (\cos (J t)+4) & -\frac{1}{16} i \sin (J t) & \cdot \\
 \cdot & \frac{1}{16} i \sin (J t) & \frac{1}{16} (4-\cos (J t)) & \cdot \\
 \cdot & \cdot & \cdot & \frac{1}{4} \\
\end{array}
\right).
\end{equation}

Due to the particular form of $\rho_{AB}$, the corresponding orbit exhibit a degeneracy which, with the stereographic parametrization, leads to the following simple form of evolution operator:

\begin{equation}
\hat{U}_{t}=\left(
\begin{array}{cccc}
 1 & \cdot & \cdot & \cdot \\
 \cdot & \frac{1}{\sqrt{1+|z(t)|^2}} &  \frac{-z^{*}(t)}{\sqrt{1+|z(t)|^2}} & \cdot \\
 \cdot & \frac{z(t)}{\sqrt{1+|z(t)|^2}} & \frac{1}{\sqrt{1+|z(t)|^2}} & \cdot \\
 \cdot & \cdot & \cdot & 1 \\
\end{array}
\right).
\end{equation}

Solving the following equation for $z(t)$:

\begin{equation}
\rho_{AB}(t)=\hat{U}_{t}\rho_{AB}(0)\hat{U}^{\dag}_{t},
\end{equation}

we obtain:

\begin{equation}
\hat{U}_{t}=\left(
\begin{array}{cccc}
 1 & \cdot & \cdot & \cdot \\
 \cdot & \frac{1}{\sqrt{1+\tan^2\left(\frac{Jt}{2}\right)}} &  \frac{i\tan\left(\frac{Jt}{2}\right)}{\sqrt{1+\tan^2\left(\frac{Jt}{2}\right)}} & \cdot \\
 \cdot &\frac{i\tan\left(\frac{Jt}{2}\right)}{\sqrt{1+\tan^2\left(\frac{Jt}{2}\right)}} & \frac{1}{\sqrt{1+\tan^2\left(\frac{Jt}{2}\right)}} & \cdot \\
 \cdot & \cdot & \cdot & 1 \\
\end{array}
\right).
\end{equation}

The bipartite Hamiltonian $H_{AB}$ in this case is time independent and reads:

\begin{equation}
H_{AB}=\left(
\begin{array}{cccc}
 0 & 0 & 0 & 0 \\
 0 & 0 & -\frac{ J}{2} & 0 \\
 0 & -\frac{ J}{2} & 0 & 0 \\
 0 & 0 & 0 & 0 \\
\end{array}
\right),
\end{equation}

which is equivalent to the following one:

\begin{equation}
H_{AB}=-\frac{J}{4}(\sigma^1_{A}\otimes\sigma^1_{B}+\sigma^2_{A}\otimes\sigma^2_{B}).
\end{equation}

In the following section we show how to extend the applicability of the stereographic parametrization method to the more general case of dissipative evolution.  

\subsection{\textit{Dissipative evolution: reconstruction of a master equation}}

Suppose that we are given a density matrix $\rho_{AB}(t)$ for our bipartite system $S = A+B$, undergoing non unitary evolution. We aim at reconstructing examples of master equations from and compatible with $\rho_{AB}(t)$. Such dissipative evolution stems from the fact that our two-qubits are interacting with another quantum system $E$. We denote the interaction term in the Hamiltonian by $H_{SE}(t)$. As stated in sec. \ref{2}, the open-system nature of time evolution reflect itself in the presence of an additional term called dissipator $\mathcal{D}_t$ in the dynamical equation for $\rho_{AB}(t)$:

\begin{equation}
\dot{\rho}_{AB}(t) = -i\left[H_{AB}(t),\rho_{AB}(t)\right]+\mathcal{D}_t\,\rho_{AB}(t).
\label{fullmaster}
\end{equation}

where, in general, $\mathcal{D}_t\,\rho_{AB}(t)=-i\textrm{Tr}_{E}\left\{[H_{SE},\rho_{SE}(t)]\right\}$. 

The approach to time evolution generators that we consider in this paper is a time convolutionless one, namely (\ref{convolutionless}):

\begin{equation}
\dot{\rho}_{AB}(t)=\mathcal{L}_{t}\rho_{AB}(t)
\end{equation}

or, in a formal relation for dynamical maps:  $\dot{\Lambda}_{t}=\mathcal{L}_{t}\Lambda_{t}$, where $\dot{\Lambda}_{t}$ denotes differentiation of the map $\Lambda_{t}$ with respect to $t$. 
An important problem in open quantum system theory is to point out physical conditions implying the  \textit{divisibility} of $\Lambda_{t}$, namely that for any $0 \leq t_1 \leq t_2$ we have \cite{NM1}: 

\begin{equation}
\Lambda_{t_2} = V_{t_2,t_1}\Lambda_{t_1}. 
\end{equation}

In particular, when the propagator of the dynamics  $V_{t_2,t_1}$ is also a CP-map, the whole dynamics is said CP-\textit{divisible} or \textit{Markovian} since it can be considered as a quantum generalization of the equation for a classical Markovian stochastic process \cite{Chru1}. Measures of the deviation of a quantum process from  Markovianity were also proposed, based on the information flow between an open system and the environment \cite{NM1}, \cite{NM2}. According to the above definition, the divisibility feature of a map is fully characterized in terms of the generator of the master equation. Indeed, the following general theorem holds \cite{alicki2}: \newline

\textbf{Theorem}: \textit{A dynamical map $\Lambda_t$ is CP-divisible (Markovian) if and only if the corresponding generator is GSKL at all times t, namely:} \newline

\begin{equation}\begin{split}
\dot{\rho}(t)&=-i\left[H(t),\rho(t)\right]+\\
&\sum_{k}\left(V_{k}(t)\rho(t) V_{k}(t)^\dag -\frac{1}{2}\{V_{k}(t)^\dag V_{k}(t),\rho(t)\}\right).
\label{GKSLt}
\end{split}\end{equation}

The first commutator term in eq. (\ref{GKSLt}) can be identified by means of the above parametrization of unitary operators. Indeed, instead of the form (\ref{BM1}) with time dependent eigenvalues, the scheme of reconstruction is applied to the following: 

\begin{equation}
\rho_{AB}(t)=\hat{u}_{t}\Gamma(t)\hat{u}_{t}^\dag,
\label{BM2}
\end{equation}

where $\Gamma(t)$ denotes the fact that the eigenvalues are not constants any more. Once we get the corresponding Hamiltonian $H_{AB}$, we are left with the following identity: 

\begin{equation}
\dot{\rho}_{AB}(t)+i\left[H_{AB},\rho_{S}(t)\right]=\mathcal{D}_t\,\rho_{AB}(t).
\label{formalidentity}
\end{equation}

Given a legitimate two qubit time evolution $\rho_{AB}(t)$ in the form (\ref{general}), the scheme of reconstruction relies on the representation of the dissipator 
as a generic affine transformation of the coherence vector $\mathbf{r}(t)$: 

\begin{equation}
\dot{\mathbf{r}}(t)-H(t)\mathbf{r}(t)=D(t)\mathbf{r}(t)+\mathbf{l}(t),
\label{affine}
\end{equation}

where $H(t)$ represents the action of the Hamiltonian term $H_{AB}(t)$ on the coherence vector. Moreover, $D(t)$ is a real $n^{2}-1 = 15$ dimensional square matrix whose entries are given by $D_{ij}(t)=\textrm{Tr}(\hat{\lambda}_{i}\Lambda_{t}[\hat{\lambda}_{j}])$ and $l_{i}(t)=\textrm{Tr}(\hat{\lambda}_{i}\Lambda_{t}\left[\mathbb{I}_{n}\right])$, where $\{\hat{\lambda}_{i}\}$ denote the 15 generators in eq.(\ref{general}).  

Eq. (\ref{affine}) can be represented in matrix form as follows: 

\begin{equation}
\left(\begin{array}{c}
1\\
\dot{\mathbf{r}}(t)-H(t)\mathbf{r}(t)
\end{array}
\right)=\left(\begin{array}{cc}
1&\vec{0}^{\textrm{\:T}}\\
\mathbf{l}(t)&D(t)
\end{array}
\right)\left(\begin{array}{c}
1\\
\mathbf{r}(t)
\end{array}
\right).
\label{matrix}
\end{equation}

Finally, the inverse problem of reconstructing the  corresponds simply to the parametrization of all possible $\mathbf{l}(t)$ and $D(t)$ such that eq. (\ref{matrix}) is satisfied for a given $\mathbf{r}(t)$. For arbitrary dimensions, this amounts at solving an undetermined system of $n^{2}-1$ equations with $n^{2}(n^{2}-1)$ unknowns. Thus, to fully characterize the dynamical map one needs the knowledge of at least $n^{2}$ time evolutions for $\rho_{AB}(t)$ \cite{buze}. 

The entire class of generators is in principle given by all possible $D(t)$, whose total superoperator $\mathcal{L}_{t}$ generates a CP evolution. 
Solving the dynamic QMP problem in full generality for only one dissipative evolution is clearly out of reach. However, let us show such difficulty by constructing at least one form of master equation from $\rho_{AB}(t)$. 

We limit the complexity by reducing the number of free parameters. A first attempt is to restrict only to \textit{unital} quantum processes, i.e. those having the maximally mixed state $\rho_{AB}=\frac{1}{4}\mathbb{I}$ as a fixed point.
This implies $\mathbf{l}(t)=\mathbf{0}$. Furthermore, we choose a symmetric form of $D(t)$ since any antisymmetric part of $D(t)$ would represent a further contribution to unitary dynamics. To understand this let us consider the lowest dimensional case, i.e. a single qubit. Let us substitute the corresponding expansion for $\rho_{S}(t)$ and $H_{S}(t)$ in terms of Pauli matrices $\sigma_{i}$ into the left side of eq.(\ref{formalidentity}), obtaining then:

\begin{equation}
i\left[H_{S},\rho_{S}(t)\right]=\frac{i}{2}\sum^{3}_{i,j=1}h^{S}_{i}(t)r_{j}(t)\left[\sigma_{i},\sigma_{j}\right].
\label{substitution}
\end{equation}

Recalling that $[\sigma_{i},\sigma_{j}]=2i\epsilon_{ijk}\sigma_{k}$, where $\epsilon_{ijk}$ is the totally antisimmetric tensor, it is easy to show that the LHS of eq. (\ref{substitution}) can be cast in the following matrix form:

\begin{equation}
2\left(\begin{array}{ccc}
0&-h^{S}_{3}(t)&h^{S}_{2}(t)\\
h^{S}_{3}(t)&0&-h^{S}_{1}(t)\\
-h^{S}_{2}(t)&h^{S}_{1}(t)&0
\end{array}\right)\left(\begin{array}{c}
{r}_{1}(t)\\
{r}_{2}(t)\\
{r}_{3}(t)
\end{array}\right).
\label{known}
\end{equation}

Thus, we recognize the skew-simmetry of the matrix representing the commutator term $i\left[H_{S},\rho_{S}(t)\right]$ acting on the Bloch vector of $\rho_{S}(t)$. Analogously, one recovers the skew-simmetry in the two-qubit case. 

Finally, we are left with the problem of finding at least one symmetric form of $D(t)$ satisfying eq. (\ref{affine}) to which correspond terms of the following GKSL representation:

\begin{equation}
\dot{\rho}_{S}=-i\left[H_{S}(t),\rho_{S}\right]+\sum^{n^{2}-1}_{i,j=1}K_{ij}(t)\left[\lambda_{i}\rho_{S} \lambda_{j}^\dag -\frac{1}{2}\{\lambda_{i}^\dag \lambda_{j},\rho_{S}\}\right],
\label{marko2}
\end{equation}

where, the matrix $K(t)=[K_{ij}(t)]$, called Kossakowski matrix, is Hermitian and the theorem above holds if and only if $K(t)$ is positive semidefinite at all $t\geq0$ \cite{BP}. 

Let us show how the entire method is applied with one example. 
Suppose, after going through a certain kinematic QMP, that we want to construct a time-local generator of two qubits whose time evolution is given as follows:

\small
\begin{equation}
\rho_{AB}(t) = \left(
\begin{array}{cccc}
 \frac{1}{2} \beta_{+}(t) \cos ^2\left(\frac{3 J t}{4}\right) & 0 & 0 & -\frac{1}{2} i  \beta_{+}(t) \sin \left(\frac{3 J t}{2}\right) \\
 0 & 0 & 0 & 0 \\
 0 & 0 & \frac{1}{2} \beta_{-}(t) & 0 \\
 \frac{1}{2} i \beta_{+}(t) \sin \left(\frac{3 J t}{2}\right) & 0 & 0 & \frac{1}{2} \beta_{+}(t) \sin ^2\left(\frac{3 J t}{4}\right) \\
\end{array}
\right)
\label{cine}
\end{equation}
\normalsize

where $\beta_{\pm}(t) =  \left(1\pm e^{-\gamma t  }\right)$. This yields the two following marginals:

\begin{equation}
\rho_{A}(t) =\frac{1}{2} \left(
\begin{array}{cc}
  \beta_{+}(t) \cos ^2\left(\frac{3 J t}{4}\right)+ \beta_{-}(t) & 0 \\
 0 &  \beta_{+}(t) \sin ^2\left(\frac{3 J t}{4}\right)\\
\end{array}
\right)
\end{equation}

\begin{equation}
\rho_{B}(t) = \frac{1}{2} \left(
\begin{array}{cc}
   \beta_{+}(t) \cos ^2\left(\frac{3 J t}{4}\right) & 0 \\
 0 &   \beta_{+}(t) \sin ^2\left(\frac{3 J t}{4}\right)+ \beta_{-}(t) \\
\end{array}
\right),
\end{equation}
\normalsize

\begin{figure}
\hspace{-0.6cm}\includegraphics[scale=0.63]{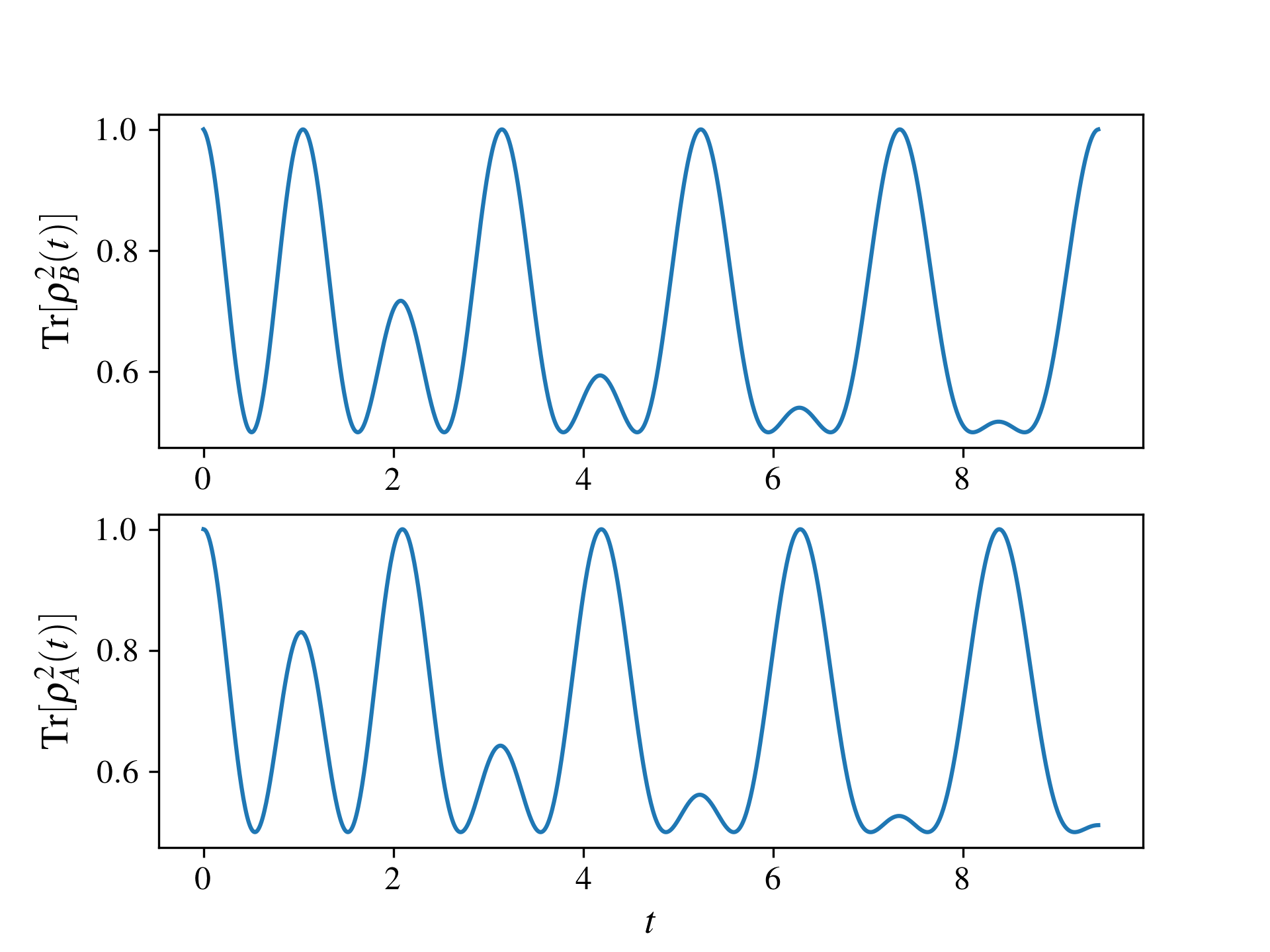}
\caption{\footnotesize Purity of marginals A and B for $J=2, \gamma=0.2$}
\label{pur}
\end{figure}

which have \textit{purities} $P(t) = \textrm{Tr}[\rho^2(t)]$ evolving as in fig.(\ref{pur}). Note that the two marginals are pure ($P=1$) at $t=0$ and then the only compatible state $\rho_{AB}(0) =\rho_A(0)\otimes\rho_B(0)$ is also pure. This simple observation hides an important implication. Indeed, the two marginals are not isospectral meaning that any compatible $\rho_{AB}(t)$ at later times has to be mixed. Therefore, from the knowledge of the marginals only, we can conclude that no unitarily evolving $\rho_{AB}(t)$ exists.   

\begin{figure}
\hspace{-0.6cm}\includegraphics[scale=0.63]{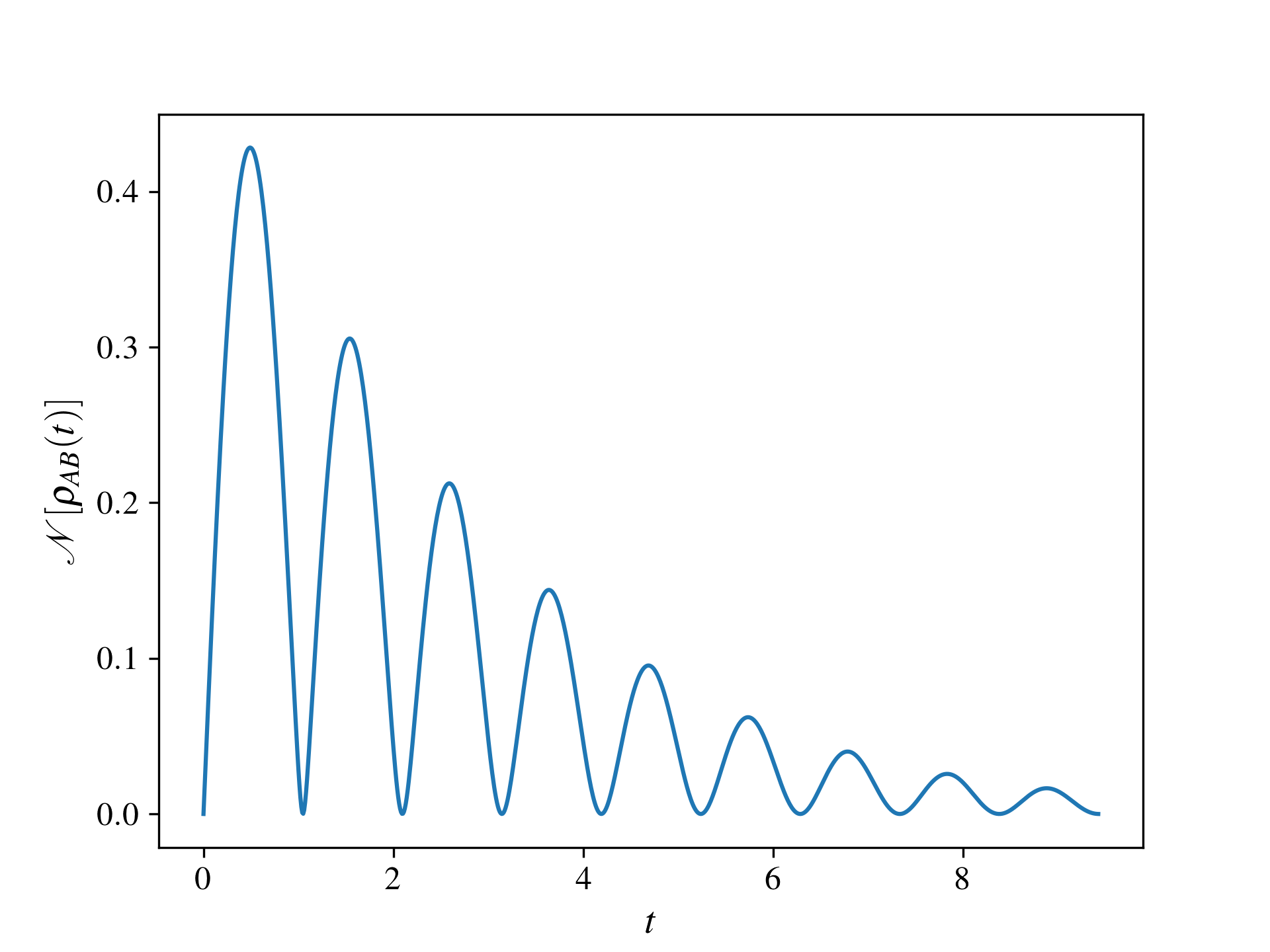}
\caption{\footnotesize Negativity of $\rho_{AB}(t)$ for $J=2, \gamma=0.2$}
\label{neg}
\end{figure}

A second relevant aspect of eq. (\ref{cine}) is the behaviour of quantum correlations developed by the systems $A$ and $B$. This can be seen analyzing the entanglement as measured by the so-called \textit{negativity}, based on the Peres-Horodecki criterion: a two-qubit state $\rho_{AB}$ is entangled iff the partially transposed state $\rho^{\tau}_{AB}=(1\otimes\tau)\rho_{AB}$ is negative definite \cite{horodecki}. The negativity is defined as follows:

\begin{equation}
\mathcal{N}(\rho_{AB}):=\frac{\left\lVert \rho^{\tau}_{AB} \right\rVert -1}{2},
\label{nega}
\end{equation} 

where $\left\lVert A \right\rVert=\textrm{Tr}\sqrt{A^{\dag}A}$ is the trace norm. It can be shown that eq. (\ref{nega}) amounts at the sum of the moduli of negative eigenvalues and thus, from eq. (\ref{cine}) we have (see fig.(\ref{neg})):

\begin{equation}
\mathcal{N}(\rho_{AB}(t)) = \frac{1}{4}\left|\,\beta_{-}(t)-\sqrt{\beta^2_{-}(t)+2e^{-2\gamma t}\left[1+\cos(3Jt)\right]}\,\right|.
\end{equation}

Applying the scheme of reconstruction discussed in the above section we have that a possible evolution operator according to eq.(\ref{BM2}) and its Hamiltonian generating the unitary part of the dynamics, are given as follows:  

\begin{equation}
\hat{U}_{t} = \left(
\begin{array}{cccc}
 \cos \left(\frac{3 J t}{4}\right) & 0 & 0 & i \sin \left(\frac{3 J t}{4}\right) \\
 0 & 1 & 0 & 0 \\
 0 & 0 & 1 & 0 \\
 i \sin \left(\frac{3 J t}{4}\right) & 0 & 0 & \cos \left(\frac{3 J t}{4}\right) \\
\end{array}
\right)
\end{equation}

\begin{equation}
H_{AB} = \left(
\begin{array}{cccc}
 0 & 0 & 0 & \frac{3 J}{4} \\
 0 & 0 & 0 & 0 \\
 0 & 0 & 0 & 0 \\
 \frac{3 J}{4} & 0 & 0 & 0 \\
\end{array}
\right)= \frac{3J}{8}(\sigma^1_{A}\otimes\sigma^1_{B}-\sigma^2_{A}\otimes\sigma^2_{B}).
\end{equation}

After a straightforward derivation, the corresponding problem of eq. (\ref{matrix}) can be derived by computing the components of the coherence vector $\mathbf{r}(t)$.
Note that possible matrix forms of $D(t)$ can be guessed much more easily from the diagonal form $\Gamma_{AB}(t)$ of $\rho_{AB}(t)$, namely:

\begin{equation}
-\gamma\left(
\begin{array}{c}
0\\
 e^{-\gamma  t}\\
  e^{-\gamma  t}
\end{array}
\right)= D'(t)\cdot \left(
\begin{array}{c}
1\\
e^{-\gamma  t}\\
 e^{-\gamma  t}
\end{array}
\right),
\end{equation}

where we have simply reduced the dimension of the problem by taking into account only the non vanishing components of $\mathbf{r}(t)$, i.e. $r_{3} = 1, r_{12}(t) = r_{15}(t) =  e^{-\gamma  t}  $. The resulting map described by $D'(t)$ has to be unitarily transformed according to the following identity:

\begin{equation}
\dot{\rho}_{AB}(t)+i\left[H_{AB},\rho_{AB}(t)\right]=\hat{U}^{\dag}_{t}\mathcal{D}'_t[\Gamma_{AB}(t)]\hat{U}_{t}.
\label{transf}
\end{equation}

As an example, one would be tempted to choose the following form of $D'(t)$ \footnote{All other elements are also imposed equal to zero.}:

\begin{equation}
D' = \textrm{diag}\left\{0, -\gamma, -\gamma\right\}. 
\label{choice1}
\end{equation}

However, such choice does not correspond to a CP dynamical map $\Lambda_t$. Indeed, for the special case of diagonal elements in $D'$ the CP conditions are easy to derive since the diagonal elements $K_{jj}$ of the Kossakowski matrix are uniquely determined as functions of ${D'_{jj}}, j=1,\dots,15$. In particular, one has that $D'$ generates a CP map if and only if $K_{jj}\geq 0$. For the choice in eq. (\ref{choice1}) one obtains the following Kossakowski matrix: 

\begin{equation}
K = \textrm{diag}\left\{0,0,-\frac{\gamma }{8},\frac{\gamma }{8},0,0,\frac{\gamma }{8},\frac{\gamma }{8},0,0,\frac{\gamma }{8},-\frac{\gamma }{8},0,0,-\frac{\gamma }{8}\right\}, 
\end{equation}

so that the presence of negative eigenvalues indicates that the above choice does not correspond to a physically legitimate scenario. One can also involve several others elements of $D'$. For instance, by choosing $D'_{ii} = 0, i=1,\dots,7$ and $D'_{jj} = -\gamma, j=8,\dots,15$ one obtains: 

\begin{equation}
K = \textrm{diag}\left\{0,0,0,\frac{\gamma }{2},0,\cdots,0\right\}.  
\end{equation}

This choice corresponds instead the following legitimate dissipator map: 

\begin{equation}
\mathcal{D'}_{t}[\rho]=\frac{\gamma}{2}\left[(\sigma_1\otimes\mathbb{I}_{2})\rho(\sigma_1\otimes\mathbb{I}_{2})-\rho\right].
\end{equation}   

Finally, one applies the unitary rotation as in eq. (\ref{transf}). 
One can in principle consider also time dependent diagonal elements for $D'(t)$ allowing then for non-divisible (non-Markovian) time evolution if such entries become negative. In such case, the CP conditions can be given in the following form:  

\begin{equation}
\Gamma_{ii}(t)=\int^{t}_{0}K_{ii}(\tau)d\tau \geq 0\quad \forall i.
\end{equation}

Interestingly, the freedom provided by only one time evolution $\rho_{AB}(t)$ implies in principle possible legitimate generators of both Markovian and Non-Markovian dynamics. Such generators can be addressed to different physical scenarios meaning that the same $\rho_{AB}(t)$ can originate from a huge variety of contexts. This difficulty can be overtaken, for example by introducing other global time evolutions from different initial conditions $\rho_{AB}(t_0)$. Such case will be considered in further studies. 

\section{Conclusions}

In this work, inspired by recent experimental progresses, we have addressed a control target in which two interacting subsystems have an
assigned dynamics of interest, reducing it to the search of a global tailored time evolution generator. Such an investigation presents several problems even in simplest case of two qubits.

The first step consists in the search of
the class of joint density matrices compatible with such time dependent marginal constraint. The time dependent nature of the problem leads us to consider it as a kinematic generalization of the quantum marginal problem.
A second time dependent quantum marginal problem arises instead naturally when we characterize a possible dynamical scenario generating a certain bipartite
evolution from an initial condition on the joint state $\rho_{AB}(t_0)$.

Our results can be regarded as an extension of the stereographic parametrization method to the case of interacting systems and non-unitary evolution. We hope that such a perspective could represent an interesting applicative platform in order to realize effective control protocols of experimental interest and to shed light on some still open problems, such as the necessary and
sufficient conditions for the legitimacy of a Non-Markovian time evolution generator.

\section{Acknowledgements}

GB and AM acknowledge M. Ku\'{s}, C. Schilling and R. Palacino for stimulating discussions.


%
%
%
%
%
%
%
%
%
%
%
%
%
%




\begin{thebibliography}{99} 
\small

\bibitem[1]{traincat}J. S. Glaser \textit{et. al}, Eur. Phys. J. D 69, 2015.

\bibitem[2]{schirm1}S. G. Schirmer, H. Fu \& A. I. Solomon, Phys. Rev. A 63.6: 063410, 2001.

\bibitem[3]{wiseman}H. M. Wiseman \& G. J. Milburn, \textit{Quantum Measurement and Control}, (Cambridge Univ. Press, Cambridge, 2009). 

\bibitem[4]{altafini}C. Altafini \& F. Ticozzi, IEEE Trans. Automat. Control, 57: 1898–1917, 2012.

\bibitem[5]{gross}M. Walter, B. Doran, D. Gross \& M. Christandl, Science 340, Issue 6137, pp. 1205-1208, 2013.

\bibitem[6]{schilling1}C. Schilling, C. L. Benavides-Riveros \& P. Vrana, Phys. Rev. A 96.5: 052312, 2017.

\bibitem[7]{mazziotti}R. Chakraborty \& D. A. Mazziotti, Phys. Rev. A 91.1: 010101, 2015.

\bibitem[8]{schirm2}S. G. Schirmer, "Hamiltonian Engineering for Quantum
Systems", In Proceedings of 3rd IFAC Workshop on Lagrangian and Hamiltonian Methods in Nonlinear Control (Nagoya, Japan 2006).

\bibitem[9]{BP}H.-P. Breuer \& F. Petruccione, \textit{The Theory of Open Quantum Systems}, (Oxford Univ. Press, Oxford, 2006).

\bibitem[10]{bernatska}J. Bernatska \& A. Messina, Physica Scripta, 85: 015001, 2012.

\bibitem[11]{bernatska2}J. Bernatska \& P. Holod, Proc. 9th Int. Conf. Geometry, Integrability and Quantization (Sofia), 146–66, 2008.



\bibitem[12]{Chru1}D. Chru\'{s}ci\'{n}ski, Open Syst. Inf. Dyn. 21: 1440004, 2014.

\bibitem[13]{rivas}A. Rivas \& S. F. Huelga, \textit{Open Quantum Systems: An Introduction}, (Springer, 2011).

\bibitem[14]{marginal}A. A. Klyachko, J. Phys. Conference Series 36: 72-86, 2006.

\bibitem[15]{schilling}C. Schilling, \textit{Quantum marginal problem and its physical relevance}, PhD thesis at ETH Zurich (2014).

\bibitem[16]{kuskus} A. Sawicki, M. Walter \& M. Ku\'{s}, J. Phys. A 46: 5, 2013.

\bibitem[17]{messina1}E. Br\"{u}ning, H. M\"{a}kel\"{a}, A. Messina, and F. Petruccione, J. Mod. Optics 59: 1, 2012.

\bibitem[18]{macdonald}I. G. Macdonald, \textit{Symmetric Functions and Hall Polynomials, Second Edition}, (Oxford Univ. Press, 1995). 


\bibitem[19]{alicki2}R. Alicki \& K. Lendi, \textit{Quantum Dynamical Semigroups and Applications}, (Springer, Berlin, 1987).


\bibitem[20]{GKS}V. Gorini, A. Kossakowski, \& E. C. G. Sudarshan, J. Math. Phys. 17: 821, 1976.

\bibitem[21]{Lindblad}G. Lindblad, Comm. Math. Phys. 48: 119, 1976.

\bibitem[22]{fano}U. Fano, Rev. Mod. Phys. 55: 855, 1983.

\bibitem[23]{kimura1}G. Kimura, Phys. Lett. A 314: 339, 2003.

\bibitem[24]{huo}S. Luo, Phys. Rev. A 77.4: 042303, 2008.

\bibitem[25]{matrix}R. A. Horn, C. R. Johnson, \textit{Matrix Analysis, Second Edition}, (Cambridge Univ. Press, Cambridge, 2013).

\bibitem[26]{hall2}B. Hall, \textit{Lie Groups, Lie Algebras, and Representations, An Elementary Introduction, Second Edition}, (Springer, Heidelberg, 2015).

\bibitem[27]{NM1}S.C. Hou, X.X Yi, S.X. Yu \& C.H. Oh, Phys. Rev. A 83.6: 062115, 2011.

\bibitem[28]{NM2}H.-P. Breuer, E.-M. Laine \& J. Piilo, Phys. Rev. Lett. 103: 210401, 2009.

\bibitem[29]{NM3}D. Chru\'{s}ci\'{n}ski, A. Kossakowski \& A. Rivas, Phys. Rev. A 83: 052128, 2011.

\bibitem[30]{buze}V. Bužek, Phys. Rev. A 58.3: 1723, 1998.

\bibitem[31]{horodecki}R. Horodecki, P. Horodecki, M. Horodecki, K. Horodecki, Rev. Mod. Phys. 81: 865, 2009.
\end{thebibliography}
\end{document}